\documentstyle[epsfig,graphicx,12pt,twoside,fleqn,espcrc1]{article}

\newcommand{\AmS}{{\protect\the\textfont2
  A\kern-.1667em\lower.5ex\hbox{M}\kern-.125emS}}

\title{A next-to-next-to-leading-order $pp \rightarrow pp\pi^0$
transition operator in chiral perturbation theory}

\author{V.~Dmitra\v{s}inovi\'{c}, K.~Kubodera, F.~Myhrer
\address{Department of Physics and Astronomy,
University of South Carolina,
Columbia, SC 29208, U. S. A.}%
and  T.~Sato,
\address{Department of Physics, Osaka University,
Toyonaka, Osaka 560-0043, Japan}}

\begin{document}
\maketitle

\begin{abstract}
We discuss the results of a systematic calculation of the
next-to-next-to-leading-order amplitude for the $pp \rightarrow pp\pi^0$ 
S-wave production at the threshold in heavy-baryon chiral perturbation theory. 
We find six new diagrams, two of which can be viewed as vertex corrections, of
15 - 20 \%, to the $\pi$-exchange graph, whereas the rest are much larger, 
one exceeding 900 \%. We discuss the reasons for this 
enhancement, as well as the steps necessary to be taken before the final
comparison with experiment.
\end{abstract}

\section{Introduction} 

Neutral pion production in proton-proton collisions $pp \rightarrow pp\pi^0$ 
near threshold needs no special introduction in the present section of these 
proceedings. The fact that the two dominant
terms in the $\pi N$ hamiltonian, the P-wave $\pi$ emission from a nucleon and 
the Weinberg-Tomozawa term are forbidden by the selection rules 
demands an exploration of 
the less-well-known subdominant terms in the hamiltonian. One of these terms goes 
by the name of the Galilean term for an S-wave pion emission operator, Fig. 1(a) 
and is common to all nonrelativistic theories, as is the isospin-symmetric part 
of the elastic $\pi N$ scattering amplitude, Fig. 1(b). 
Since the exchanged $\pi^0$ in Fig. 1(b) is off-shell, this graph is sensitive 
to the off-shell extrapolation of the 
$\pi N$ amplitude, which is not unique, however.
All terms subdominant to the ones above are model-dependent. 
Precise cross section data measured in Bloomington 
\cite{blooming} and Uppsala \cite{uppsala} are roughly five times larger 
\cite{lr93} 
than the simplest widely agreed-on prediction of the two diagrams in Fig. 1. 

%
\begin{figure}[htb]
\begin{minipage}[t]{155mm}
\framebox[0mm]{\rule[0mm]{0mm}{0mm}}
\includegraphics*[bb= -5 640 317 750,scale=1.1]  
{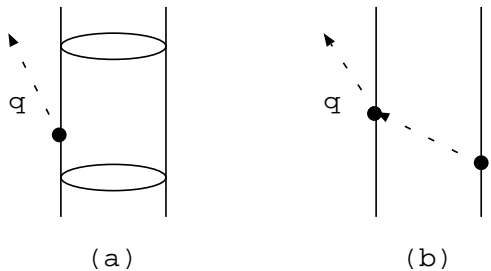}
\caption{The Born approximation (a) and the $\pi$ rescattering graph (b) in 
$pp \rightarrow pp\pi^0$.}
\label{fig:toobig}
\end{minipage}
\end{figure}
One systematic proposal for corrections to graphs in Fig. 1 is based on the
underlying chiral symmetry of the strong interactions and goes by the name 
of heavy-baryon chiral perturbation theory [HB$\chi$PT]. 
HB$\chi$PT generates additional subdominant terms in the hamiltonian that are
clearly ordered by means of so-called chiral counting. 

This work is subject to the following assumptions:
{\it 1. Heavy-baryon chiral perturbation theory} This means that we include: 
(i) only pion and nucleon degrees of freedom, i.e., no mesons heavier than the 
pion and no nucleon resonances; (ii) static nucleons in loop integrals and
non-relativistic external nucleons, i.e., one can use the theory only near the 
threshold; (iii) well defined chiral order counting; this theory ought to 
provide smaller and smaller corrections to the leading-order result; (iv) 
perturbation theory, i.e., a finite number of graphs and counterterms, hence
no dynamically created resonances or bound states.
{\it 2. Plane-wave initial and final states} This means that
we do not include the distortions of the initial and final waves, 
but give a brief discussion of the problems involved and of 
our present efforts to calculate the distorted wave effects.

\section{Results}

We present the results of a systematic analysis of 
next-to-next-to-leading-order
amplitude for the $pp \rightarrow pp\pi^0$ production 
at the threshold in heavy-baryon chiral perturbation theory. 
We found 19 topologically distinct new types of one-loop diagrams and one $\pi$ 
rescattering correction, that can potentially contribute to this reaction. 
The isospin selection rules and the S-wave character of the outgoing pion
reduce the number of graphs from 19 to 6 (shown in Fig. 2).
\begin{figure}[htb]
\begin{minipage}[t]{160mm}
\framebox[0mm]{\rule[0mm]{0mm}{0mm}}
\includegraphics*[bb= -15 385 527 750,scale=0.7]  
{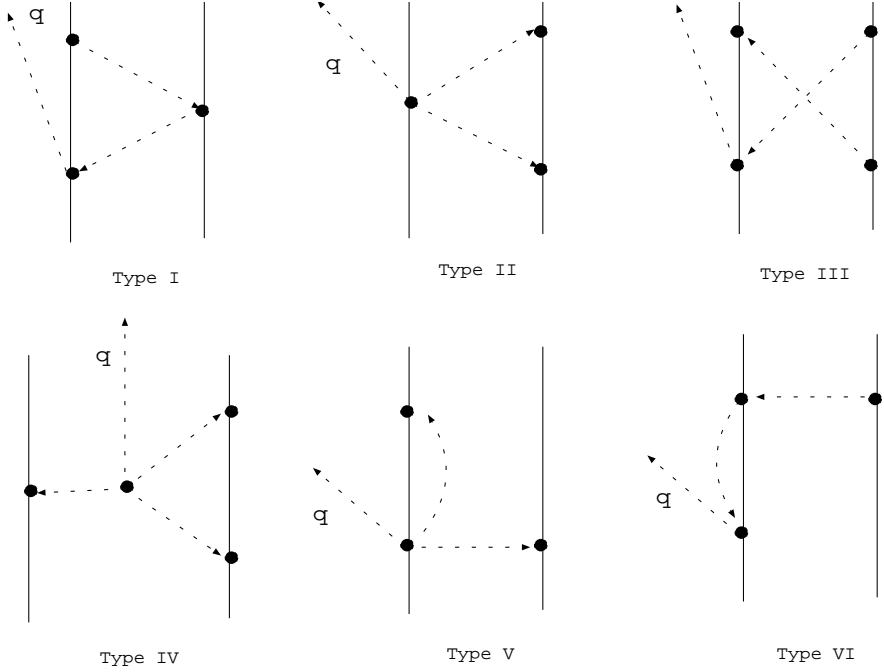}
\caption{The six types of one-loop next-to-next-to-leading order diagrams. 
The ${\cal O}(M^{-1})$ ``recoil'' correction to $\pi$ rescattering graph, 
type VII, is topologically equivalent to Fig. 1.(b).}
\label{fig:toobig3}
\end{minipage}
\end{figure}
Analytic expressions for these effective transition operators are given in 
Ref. \cite{dkms99}.
In Table I we present our results on the relative importance of the six new 
types of diagrams, as well as the pion rescattering correction (type VII) that
is topologically equivalent to the graph in Fig 1(b),
as compared with the one-pion exchange rescattering graph in the
plane-wave approximation for the initial and final states.
\begin{table}[hbt]
\caption{
Sizes of the $K^{\rm th}$ type of diagrams
relative to the $\nu$ = 1 pion rescattering diagram, Fig. 1(b). 
The ratio $R_K$ defined in the text is given for 
three sets (A, B and C) of parameters' $c_{1,2,3}$ values, 
see the text and Table 1 in Ref. \cite{dkms99}.}
\label{tab:horizontal}
\begin{tabular*}{\textwidth}{@{}l@{\extracolsep{\fill}}rrrrrrr}
\hline
{$K$} & {I} & {II} & {III} & {IV} & {V} & {VI} & {VII} \\
\hline
{${\rm A}$} & {$- 0.70$} & {$6.7$} & {$-6.7$} & {$9.5$} & {$0.18$} & {$0.14$} &
{$2.6$} \\
{${\rm B}$} & {$- 0.38$} & {$3.6$} & {$-3.6$} & {$5.1$} & {$0.10$} & {$0.08$} &
{$1.9$}  \\
{${\rm C}$} & {$- 0.53$} & {$5.1$} & {$-5.1$} & {$7.2$} & {$0.14$} & {$0.11$} &
{$2.0$} \\ 
\hline
\end{tabular*}
\end{table}
The vertex-correction types of graphs, V and VI, are found to give only 
small corrections to the lower order result of Fig. 1(b), in conformity with 
expectations from general tenets of chiral perturbation theory.
By contrast, we find very large contributions from the two-pion exchange 
graphs I - IV. Two of these types of graphs, II and III, cancel to within
1 \% at the threshold, though they have very different energy-
and momentum transfer dependencies at other kinematics, a point of some 
importance in distorted wave calculations.
These ``genuine'' one-loop graphs can perhaps be interpreted as a part of an
``effective $\sigma$-meson exchange'' that is a significant part of an 
alternative theory based on the suggestion that the subdominant terms 
are induced by heavier-meson exchanges \cite{lr93}. The $\pi$ rescattering 
correction, type VII, is also large.

\section{Discussion}

Appearance of large individual contributions calls for an explanation.
Enhancement of the $\pi$ rescattering diagram VII over the leading-order
graph Fig.1.b is perhaps the easiest to understand: The latter
is proportional to the energy transfer between the protons $k_{0}$ which equals 
$m_{\pi}/2$ at the threshold, whereas the former is proportional to the 
three-momentum transfer $\sqrt{{\vec k}^2}$, which is determined by the kinematics 
of this reaction as 
$\sqrt{{\vec k}^2} \simeq \sqrt{m_{\pi} m_{N}} \simeq \sqrt{7} m_{\pi}$.
 
The one-loop two-pion-exchange diagrams types I - IV, on the other hand,
all involve one-loop integrals with two pion (and one nucleon) propagators
\begin{eqnarray}
I_{(\mu; \alpha,\beta)} (\omega, v , P) & = &
\frac{1}{i}\int \frac{d^4 l}{(2\pi)^4}
\frac{(l_{\mu};l_{\alpha}l_{\beta})}{( v\!\cdot\! l - \omega -i\varepsilon)
(m_\pi^2 - l^2 - i\varepsilon)
\left[m_\pi^2-(l-P)^2 - i\varepsilon \right]}
\label{e:io} 
\end{eqnarray} 
In these loop integrals $P$ typically corresponds to the four-momentum transfer 
$k$ between the protons, which 
brings a factor $k^2 = - m_{\pi} m_{N}$ into the numerator accompanied 
by $f_{\pi}^{-2}$ coming from the vertices,
thus resulting in a large enhancement factor,
${\cal O}(k^2 f_{\pi}^{-2} = - m_{\pi} m_{N}  f_{\pi}^{-2} \simeq - 15)$. 
This explains the large size of the two-pion exchange diagrams,
although their precise size is, of course, specific to the diagram. 

The deeper reason for the enhancement of both the loop graphs I - IV, and the
rescattering graph VII lies in the derivative nature
of the pion-nucleon couplings in $\chi$PT: it is the derivative coupling that
puts the loop momentum $l_{\mu}$ into the numerator of the integral (\ref{e:io}) 
and, as shown above, and it is the derivative coupling that makes 
the graph VII grow with the increasing three-momentum exchange ${\vec k}$.


\section{Status of the distorted wave calculation}

It is known \cite{slmk97} that the rescattering diagram Fig. 1(b) is 
sufficiently large to describe the data by itself, but
it also interferes with the leading Born diagram Fig. 1(a).
The latter gives a non-zero contribution only when evaluated between distorted
waves, as four-momentum conservation prevents spontaneous pion emission. 
Consequently initial and final wave distortions are crucial to any numerical 
predictions of this reaction's cross section.

Preliminary work on the distorted-wave amplitude has led us to the following
conclusions: (1) the $\pi^0$ production operator
is not square integrable: there are contact 
terms proportional to Dirac delta functions in configuration space. 
Consequently, the Fourier transforms from configuration to momentum space and
{\it vice versa} are ill-defined;
(2) the momentum space distorted wave integrals of the
$\pi^0$ production operator receive significant contributions from
momenta up to 2 GeV, far beyond the region of applicability of HB$\chi$PT;
(3) combined use of a $\pi^0$ production operator
that requires both energy and momentum transfer and two-proton 
Schr\" odinger wave functions that allow only three-momentum transfer 
can lead to inconsistencies.

We wish to thank Shung-ichi Ando for pointing out an error in our
calculation of the rescattering diagram, type VII.


\begin{thebibliography}{9}
\bibitem{blooming}
H. O. Meyer {\it et al.\/},
Phys. Rev. Lett. {\bf 65} (1990) 2846;
Nucl. Phys. {\bf A539} (1992) 633.

\bibitem{uppsala} A. Bondar {\it et al.\/},
Phys. Lett. B {\bf 356} (1995) 8.

\bibitem{lr93} see
C. Hanhart, and K. Tamura, these proceedings; for older literature see 
T.-S. H. Lee and D. O. Riska, Phys. Rev. Lett. {\bf 70} (1993) 2237. 



\bibitem{dkms99} 
V.~Dmitra\v{s}inovi\'{c}, K.~Kubodera, F.~Myhrer and T. Sato, 
LANL bulletin board e-print nucl-th/9902048.

\bibitem{slmk97}
T. Sato, T.-S. Lee, F. Myhrer and K. Kubodera, 
Phys. Rev. C {\bf 56}, 1246 (1997). 






\end{thebibliography}
\end{document}